\documentclass[aps,prd,twocolumn,superscriptaddress,showpacs]{revtex4}

\usepackage{float}
\usepackage{eso-pic} 
\usepackage{setspace}
\usepackage{color}

 \usepackage{graphicx}                                  
 \DeclareGraphicsExtensions{.eps,.ps,.eps.gz,.ps.gz}

\def\urltilde{\kern -.15em\lower .7ex\hbox{\~{}}\kern .04em}
\def\urldot{\kern -.10em.\kern -.10em}
\def\urlhttp{http\kern -.10em\lower -.1ex\hbox{:}\kern -.12em\lower 0ex\hbox{/}\kern -.18em\lower 0ex\hbox{/}}

\def\urltilde{\kern -.15em\lower .7ex\hbox{\~{}}\kern .04em}
\def\urldot{\kern -.10em.\kern -.10em}
\def\urlhttp{http\kern -.10em\lower -.1ex\hbox{:}\kern -.12em\lower 0ex\hbox{/}\kern -.18em\lower 0ex\hbox{/}}

\def\mevc {\ifmmode {\rm MeV}/c \else MeV$/c$\fi}
\def\mevcc {\ifmmode {\rm MeV}/c^2 \else MeV$/c^2$\fi}
\def\gevc {\ifmmode {\rm GeV}/c \else GeV$/c$\fi}
\def\gevcc {\ifmmode {\rm GeV}/c^2 \else GeV$/c^2$\fi}
\def\tevcc {\ifmmode {\rm TeV}/c^2 \else TeV$/c^2$\fi}

\def\ol   {\overline}

\def\vtd  {\ifmmode |V_{td}| \else $|V_{td}|$\fi}
\def\vtb  {\ifmmode |V_{tb}| \else $|V_{tb}|$\fi}
\def\vts  {\ifmmode |V_{ts}| \else $|V_{ts}|$\fi}
\def\vcb  {\ifmmode |V_{cb}| \else $|V_{cb}|$\fi}

\newcommand{\Ds} {\ifmmode D_{\mbox{\sl s}}^{-}
                       \else $D_{\mbox{\sl s}}^{-}$\fi}
\newcommand{\Bs} {\ifmmode B_{\mbox{\sl s}}^{0}
                       \else $B_{\mbox{\sl s}}^{0}$\fi}
\newcommand{\Bsb} {\ifmmode \ol B_{\mbox{\sl s}}^{0}
                       \else $\ol B_{\mbox{\sl s}}^{0}$\fi}
\newcommand{\Bsh} {\ifmmode B_{\mbox{\sl s}}^H
                       \else $B_{\mbox{\sl s}}^H$\fi}
\newcommand{\Bsl} {\ifmmode B_{\mbox{\sl s}}^L
                       \else $B_{\mbox{\sl s}}^L$\fi}
\newcommand{\Dsl} {\ifmmode D_{\mbox{\sl s}}^{-} \ell^+
                       \else $D_{\mbox{\sl s}}^{-} \ell^+$\fi}
\newcommand{\xs} {\ifmmode x_{\mbox{\sl s}}
                       \else $x_{\mbox{\sl s}}$\fi}
\newcommand{\xd} {\ifmmode x_d \else $x_d$\fi}
\newcommand{\lxy} {\ifmmode L_{\rm xy} \else $L_{\rm xy}$\fi}
\newcommand{\dgam} {\ifmmode \Delta\Gamma \else $\Delta\Gamma$\fi}
\newcommand{\dm} {\ifmmode \Delta m \else $\Delta m$\fi}
\newcommand{\ctau} {\ifmmode c\tau \else $c\tau$\fi}

\begin{document}

\thispagestyle{empty}

\vspace*{-2cm}

\vfil

\title{Measurement of $t\bar{t}$ Spin Correlation in $p\bar{p}$ Collisions Using the CDF~II Detector at the Tevatron}

\affiliation{Institute of Physics, Academia Sinica, Taipei, Taiwan 11529, Republic of China} 
\affiliation{Argonne National Laboratory, Argonne, Illinois 60439, USA} 
\affiliation{University of Athens, 157 71 Athens, Greece} 
\affiliation{Institut de Fisica d'Altes Energies, Universitat Autonoma de Barcelona, E-08193, Bellaterra (Barcelona), Spain} 
\affiliation{Baylor University, Waco, Texas 76798, USA} 
\affiliation{Istituto Nazionale di Fisica Nucleare Bologna, $^z$University of Bologna, I-40127 Bologna, Italy} 
\affiliation{University of California, Davis, Davis, California 95616, USA} 
\affiliation{University of California, Los Angeles, Los Angeles, California 90024, USA} 
\affiliation{Instituto de Fisica de Cantabria, CSIC-University of Cantabria, 39005 Santander, Spain} 
\affiliation{Carnegie Mellon University, Pittsburgh, Pennsylvania 15213, USA} 
\affiliation{Enrico Fermi Institute, University of Chicago, Chicago, Illinois 60637, USA}
\affiliation{Comenius University, 842 48 Bratislava, Slovakia; Institute of Experimental Physics, 040 01 Kosice, Slovakia} 
\affiliation{Joint Institute for Nuclear Research, RU-141980 Dubna, Russia} 
\affiliation{Duke University, Durham, North Carolina 27708, USA} 
\affiliation{Fermi National Accelerator Laboratory, Batavia, Illinois 60510, USA} 
\affiliation{University of Florida, Gainesville, Florida 32611, USA} 
\affiliation{Laboratori Nazionali di Frascati, Istituto Nazionale di Fisica Nucleare, I-00044 Frascati, Italy} 
\affiliation{University of Geneva, CH-1211 Geneva 4, Switzerland} 
\affiliation{Glasgow University, Glasgow G12 8QQ, United Kingdom} 
\affiliation{Harvard University, Cambridge, Massachusetts 02138, USA} 
\affiliation{Division of High Energy Physics, Department of Physics, University of Helsinki and Helsinki Institute of Physics, FIN-00014, Helsinki, Finland} 
\affiliation{University of Illinois, Urbana, Illinois 61801, USA} 
\affiliation{The Johns Hopkins University, Baltimore, Maryland 21218, USA} 
\affiliation{Institut f\"{u}r Experimentelle Kernphysik, Karlsruhe Institute of Technology, D-76131 Karlsruhe, Germany} 
\affiliation{Center for High Energy Physics: Kyungpook National University, Daegu 702-701, Korea; Seoul National University, Seoul 151-742, Korea; Sungkyunkwan University, Suwon 440-746, Korea; Korea Institute of Science and Technology Information, Daejeon 305-806, Korea; Chonnam National University, Gwangju 500-757, Korea; Chonbuk National University, Jeonju 561-756, Korea} 
\affiliation{Ernest Orlando Lawrence Berkeley National Laboratory, Berkeley, California 94720, USA} 
\affiliation{University of Liverpool, Liverpool L69 7ZE, United Kingdom} 
\affiliation{University College London, London WC1E 6BT, United Kingdom} 
\affiliation{Centro de Investigaciones Energeticas Medioambientales y Tecnologicas, E-28040 Madrid, Spain} 
\affiliation{Massachusetts Institute of Technology, Cambridge, Massachusetts 02139, USA} 
\affiliation{Institute of Particle Physics: McGill University, Montr\'{e}al, Qu\'{e}bec, Canada H3A~2T8; Simon Fraser University, Burnaby, British Columbia, Canada V5A~1S6; University of Toronto, Toronto, Ontario, Canada M5S~1A7; and TRIUMF, Vancouver, British Columbia, Canada V6T~2A3} 
\affiliation{University of Michigan, Ann Arbor, Michigan 48109, USA} 
\affiliation{Michigan State University, East Lansing, Michigan 48824, USA}
\affiliation{Institution for Theoretical and Experimental Physics, ITEP, Moscow 117259, Russia}
\affiliation{University of New Mexico, Albuquerque, New Mexico 87131, USA} 
\affiliation{Northwestern University, Evanston, Illinois 60208, USA} 
\affiliation{The Ohio State University, Columbus, Ohio 43210, USA} 
\affiliation{Okayama University, Okayama 700-8530, Japan} 
\affiliation{Osaka City University, Osaka 588, Japan} 
\affiliation{University of Oxford, Oxford OX1 3RH, United Kingdom} 
\affiliation{Istituto Nazionale di Fisica Nucleare, Sezione di Padova-Trento, $^{aa}$University of Padova, I-35131 Padova, Italy} 
\affiliation{LPNHE, Universite Pierre et Marie Curie/IN2P3-CNRS, UMR7585, Paris, F-75252 France} 
\affiliation{University of Pennsylvania, Philadelphia, Pennsylvania 19104, USA}
\affiliation{Istituto Nazionale di Fisica Nucleare Pisa, $^{bb}$University of Pisa, $^{cc}$University of Siena and $^{dd}$Scuola Normale Superiore, I-56127 Pisa, Italy} 
\affiliation{University of Pittsburgh, Pittsburgh, Pennsylvania 15260, USA} 
\affiliation{Purdue University, West Lafayette, Indiana 47907, USA} 
\affiliation{University of Rochester, Rochester, New York 14627, USA} 
\affiliation{The Rockefeller University, New York, New York 10065, USA} 
\affiliation{Istituto Nazionale di Fisica Nucleare, Sezione di Roma 1, $^{ee}$Sapienza Universit\`{a} di Roma, I-00185 Roma, Italy} 

\affiliation{Rutgers University, Piscataway, New Jersey 08855, USA} 
\affiliation{Texas A\&M University, College Station, Texas 77843, USA} 
\affiliation{Istituto Nazionale di Fisica Nucleare Trieste/Udine, I-34100 Trieste, $^{ff}$University of Trieste/Udine, I-33100 Udine, Italy} 
\affiliation{University of Tsukuba, Tsukuba, Ibaraki 305, Japan} 
\affiliation{Tufts University, Medford, Massachusetts 02155, USA} 
\affiliation{University of Virginia, Charlottesville, Virginia  22906, USA}
\affiliation{Waseda University, Tokyo 169, Japan} 
\affiliation{Wayne State University, Detroit, Michigan 48201, USA} 
\affiliation{University of Wisconsin, Madison, Wisconsin 53706, USA} 
\affiliation{Yale University, New Haven, Connecticut 06520, USA} 
\author{T.~Aaltonen}
\affiliation{Division of High Energy Physics, Department of Physics, University of Helsinki and Helsinki Institute of Physics, FIN-00014, Helsinki, Finland}
\author{B.~\'{A}lvarez~Gonz\'{a}lez$^v$}
\affiliation{Instituto de Fisica de Cantabria, CSIC-University of Cantabria, 39005 Santander, Spain}
\author{S.~Amerio}
\affiliation{Istituto Nazionale di Fisica Nucleare, Sezione di Padova-Trento, $^{aa}$University of Padova, I-35131 Padova, Italy} 

\author{D.~Amidei}
\affiliation{University of Michigan, Ann Arbor, Michigan 48109, USA}
\author{A.~Anastassov}
\affiliation{Northwestern University, Evanston, Illinois 60208, USA}
\author{A.~Annovi}
\affiliation{Laboratori Nazionali di Frascati, Istituto Nazionale di Fisica Nucleare, I-00044 Frascati, Italy}
\author{J.~Antos}
\affiliation{Comenius University, 842 48 Bratislava, Slovakia; Institute of Experimental Physics, 040 01 Kosice, Slovakia}
\author{G.~Apollinari}
\affiliation{Fermi National Accelerator Laboratory, Batavia, Illinois 60510, USA}
\author{J.A.~Appel}
\affiliation{Fermi National Accelerator Laboratory, Batavia, Illinois 60510, USA}
\author{A.~Apresyan}
\affiliation{Purdue University, West Lafayette, Indiana 47907, USA}
% ...
\collaboration{CDF Collaboration\footnote{With visitors from $^a$University of Massachusetts Amherst, Amherst, Massachusetts 01003, USA,
$^b$Istituto Nazionale di Fisica Nucleare, Sezione di Cagliari, 09042 Monserrato (Cagliari), Italy,
$^c$University of California Irvine, Irvine, CA  92697, USA, 
$^d$University of California Santa Barbara, Santa Barbara, CA 93106, USA,
$^e$University of California Santa Cruz, Santa Cruz, CA  95064, USA,
$^f$CERN,CH-1211 Geneva, Switzerland,
$^g$Cornell University, Ithaca, NY  14853, USA,
$^h$University of Cyprus, Nicosia CY-1678, Cyprus, 
$^i$University College Dublin, Dublin 4, Ireland,
$^j$University of Fukui, Fukui City, Fukui Prefecture, Japan 910-0017,
$^k$Universidad Iberoamericana, Mexico D.F., Mexico,
$^l$Iowa State University, Ames, IA  50011, USA,
$^m$University of Iowa, IA City, IA  52242, USA,
$^n$Kinki University, Higashi-Osaka City, Japan 577-8502,
$^o$Kansas State University, Manhattan, KS 66506, USA,
$^p$University of Manchester, Manchester M13 9PL, England,
$^q$Queen Mary, University of London, London, E1 4NS, England,
$^r$Muons, Inc., Batavia, IL 60510, USA,
$^s$Nagasaki Institute of Applied Science, Nagasaki, Japan, 
$^t$National Research Nuclear University, Moscow, Russia,
$^u$University of Notre Dame, Notre Dame, IN 46556, USA,
$^v$Universidad de Oviedo, E-33007 Oviedo, Spain, 
$^w$Texas Tech University, Lubbock, TX  79609, USA,
$^x$Universidad Tecnica Federico Santa Maria, 110v Valparaiso, Chile,
$^y$Yarmouk University, Irbid 211-63, Jordan,
$^{gg}$On leave from J.~Stefan Institute, Ljubljana, Slovenia, 
}}
\noaffiliation

\begin{abstract}
The $t\bar{t}$ spin correlation at production is a fundamental prediction of QCD and a potentially incisive test of new physics coupled to top quarks. We measure the $t\bar{t}$ spin state in $p\bar{p}$ collisions at $\sqrt{s} = 1.96$ TeV using 1001 candidate events in the lepton plus jets decay channel reconstructed in the CDF~II detector. In the helicity basis, for a top-quark mass of 172.5 GeV/$c^2$, we find a spin correlation coefficient $\kappa$ = 0.60 $\pm$ 0.50 (stat) $\pm$ 0.16 (syst), consistent with the QCD prediction, $\kappa \approx 0.40$.

\end{abstract}

\maketitle

In quark-pair production by the strong interaction, the quark spins are entangled according to the short distance dynamics of quantum chromodynamics (QCD)~\cite{willen}. The spin state is observable in angular correlations among the quark decay products induced by the V-A (Vector minus Axial-vector) nature of the weak interaction but is typically destroyed by the depolarizing effects of hadronization before the decay can proceed. The top quark is an exception to this rule. Because of its large mass, the top-quark lifetime is shorter than the fragmentation timescale, cutting off the long distance QCD effects and transmitting the $t \bar{t}$ production configuration to the final state. Measurement of the $t \bar{t}$ spin configuration is a first look at a bare-quark pair at production. The measurement tests the fundamental predictions of QCD~\cite{willen,theory1,Mahlon_Parke_paper,theorydl,theory2} and could be a sensitive discriminant of new physics coupled to top quarks~\cite{nlo,kkgrav}.  For example, a $t\bar{t}$ resonance appearing as an excess in the $t\bar{t}$ invariant-mass spectrum can be verified as a Kaluza-Klein graviton through measurement of the spin correlation as described in Ref.~\cite{kkgrav}.

Because final state charged leptons have the strongest correlation to the top-quark spin, the $t\bar{t}$ spin correlation is usually discussed in terms of the dilepton final state $t\bar{t} \rightarrow (W^+b)(W^-\bar{b}) \rightarrow (\bar{\ell} \nu)(\ell^{\prime} \bar{\nu^{\prime}}) b\bar{b}$~\cite{theorydl}. This mode suffers from a small branching ratio and poor definition of the top-quark kinematics due to the presence of two undetectable neutrinos. A previous measurement of the $t \bar{t}$ spin correlation was limited to a small sample of just six events in this mode~\cite{D0paper}. 

We report on a new measurement of the $t \bar{t}$ spin correlation in $p \bar{p}$ collisions at 1.96 TeV with a data sample corresponding to an integrated luminosity of 4.3 fb$^{-1}$ collected with the CDF~II detector at the Fermilab Tevatron. We measure the spin correlation of pair-produced quarks for the first time in the lepton plus jets decay topology, $t\bar{t} \rightarrow (W^+b)(W^-\bar{b}) \rightarrow (u \bar{d} b)(\ell \bar{\nu} \bar{b})$ or $t\bar{t} \rightarrow (W^+b)(W^-\bar{b}) \rightarrow (\bar{\ell} \nu b)(\bar{u} d \bar{b})$~\cite{particledef}.  In this decay mode, we take advantage of a large branching ratio compared to the dilepton channel and the well-constrained $t \bar{t}$ kinematics in the lepton plus jets final state with only one neutrino. The measurement relies critically on a new technique for identifying the final state down-type quark ($d$ or $s$), which has the same spin-analyzing power as a charged lepton. We expect the spin correlation measurement to show the dominance of $t\bar{t}$ production via the J=1 $q\bar{q}$ annihilation channel that occurs in $\sim$85\% of $p\bar{p}$ collisions at the Tevatron~\cite{qq_frac}.  

We work in the helicity basis, where the spin-quantization axis is defined as the direction of motion of the $t$ (or $\bar{t}$) quark in the $t\bar{t}$ rest frame.  There are other quantization axes which predict a larger value for the spin correlation~\cite{Mahlon_Parke_paper}, but they do not provide any significant increase in the statistical sensitivity of our approach, so we work with the simpler helicity basis.  A quark is called right-handed ($t_R$)/left-handed ($t_L$) if its spin is oriented along/opposite to its direction of motion.  In the $t\bar{t}$ rest frame the quarks move back-to-back; thus the same-spin states with J=1 are those with opposite helicity: $\bar{t}_L t_{R}$ and $\bar{t}_R t_{L}$. Near the energy threshold for $t \bar{t}$ production, the opposite-helicity fraction is predicted in the standard model (SM) to be $\sim$67\% for $t\bar{t}$ production via $q\bar{q}$ annihilation, while for top quarks with large momenta compared to the top-quark mass, helicity is approximately conserved and this fraction rises to $\sim$100\%~\cite{willen, Mahlon_Parke_paper}. Integrating over all top-quark momenta according to the parton distribution functions and adding the small ($\sim$15\%) $J=0$ contribution from gluon-gluon fusion processes, we expect to find an opposite-helicity fraction~\cite{willen,Mahlon_Parke_paper}
\begin{eqnarray}
F_{OH}& =&\frac{\sigma(\bar{t}_R t_L)+\sigma(\bar{t}_L t_R)}{\sigma(\bar{t}_R t_R)+\sigma(\bar{t}_L t_L)+\sigma(\bar{t}_R t_L)+\sigma(\bar{t}_L t_R)}\nonumber \\
      &\approx&0.70. 
\end{eqnarray}

$F_{OH}$ is simply related to the spin correlation coefficient $\kappa$ that measures the fractional difference between the number of events in which the top-quark spins are aligned and the number of events in which they have opposite directions: $\kappa = 2F_{OH}-1$.  We thus expect $\kappa \approx 0.40$~\cite{willen,Mahlon_Parke_paper}, while for uncorrelated spins, $\kappa = 0.0$ and $F_{OH} = 0.5$. 

In top-quark decays in the SM the V-A couplings fix the angular distributions of the decay products according to the polarization of the parent top quark via 

\begin{equation}
  \label{decay_correlation}
  \frac{1}{\Gamma} \frac{d \Gamma}{d \cos \theta_{i}} = \frac{1}{2} (1 \pm A_{i} \cos \theta_{i}),
\end{equation}

\noindent where the positive/negative sign is used for right-handed/left-handed quarks, and the helicity angle $\theta_{i}$ is defined as the angle between the spin-quantization direction and the momentum of the decay particle in the rest frame of its parent top quark.  In the V-A weak decay, the spin-analyzing-power coefficient $A_i$ is equal to +1.0 for the charged lepton or down-type quark, -0.41 for the bottom quark, and -0.31 for the neutrino or up-type quark, with the signs reversed for antitop-quark decays~\cite{Mahlon_Parke_paper}. The $t\bar{t}$ spin correlation connects the daughter helicity angles on each side of the decay.  The differential cross-section in these variables is

\begin{equation}
  \label{dsigma_correlation}
  \frac{1}{\sigma}\frac{d^2\sigma}{d(\cos\theta_i)d(\cos\theta_j)}=\frac{1+\kappa A_i A_j \cos\theta_i \cos\theta_j}{4},
\end{equation}

\noindent where $i$ and $j$ refer to top-quark and antitop-quark decay products respectively~\cite{Mahlon_Parke_paper}.

For each of the four possible $t\bar{t}$ helicity states, we create model templates for the distributions of $\cos\theta_l\cos\theta_d$ and  $\cos\theta_l\cos\theta_b$, where the charged lepton $l$ is a decay product from one top quark in the pair and the quarks $d$ and $b$ are decay products from the other quark.  We then find the relative normalization of these model templates that gives the best fit to a two-dimensional distribution of these variables in the data.  The model templates account for all acceptance effects and dilutions due to event reconstruction, so that the parton-level value of $F_{OH}$ follows directly from the template fit to the data.

CDF~II~\cite{cdfdet} is a general purpose, azimuthally and forward-backward symmetric detector.  Charged-particle directions and momenta are measured with a silicon tracker~\cite{cdfsi} and a drift chamber~\cite{cdfcot} in a 1.4 T solenoidal magnetic field.  Electromagnetic and hadronic calorimeters~\cite{cdfcal} are located beyond the solenoid and allow for jet and missing $E_T$ reconstruction.  Beyond the calorimeter, muon chambers~\cite{cdfmuon} provide coverage for the pseudorapidity range $|\eta| \le 1.0$. We use a cylindrical coordinate system with its origin at the center of the detector and the $z$ axis along the proton direction~\cite{coords}.

Lepton plus jets events are selected by requiring one electron or muon with transverse momentum of at least 20 GeV/$c$ and $|\eta|<1.0$, missing transverse energy of at least 20 GeV, and four or more jets with transverse energy of at least 20 GeV and $|\eta|<2.0$, at least one of which must be tagged as a $b$ jet by the presence of a displaced secondary vertex~\cite{btagging2}. This selection yields 1001 total candidate events, 224 of which have two tagged $b$ jets.

Non-$t\bar{t}$ backgrounds are well-constrained by precision $t\bar{t}$ cross-section measurements~\cite{M24U}, with a predicted total of 215 $\pm$ 48 background events. Non-$t\bar{t}$ models are checked against background-enriched sidebands with no tagged $b$ jets and are found to give very good representations of the normalizations and kinematics in all variables, including lepton and jet energies and angular distributions.

The helicity angles are determined in a complete reconstruction of the $t\bar{t}$ kinematics in $t\bar{t} \rightarrow (Wb)(Wb) \rightarrow (\ell \nu b)(udb)$, where we constrain $M(\ell \nu) = M(ud) = 80.4$ GeV/$c^2$, the mass of the W boson, and $M(\ell \nu b) = M(udb) = 172.5$ GeV/$c^2$, the top-quark mass, and require any tagged $b$ jets to be identified with $b$ partons.  The constraints were chosen to be close to the world averages in Ref.~\cite{PDG}.  Each of the 24 possible jet-to-parton assignments is evaluated using a $\chi^2$ comparison to the $t\bar{t}$ hypothesis with the above constraints, and we choose the assignment with the lowest $\chi^2$ value~\cite{topmassreco}. This procedure correctly assigns all jets to the corresponding partons in approximately 37\% of events.  All effects of angular acceptance and jet reconstruction and misassignment are fully modeled by our simulated samples.

Down-type-quark identification relies on the V-A decay correlation that tends to send the down-type quark in the direction opposite that of the hadronically decaying W boson in the top-quark rest frame.  We therefore assign the down-type quark as the jet that, in the W boson rest frame, is closest to the bottom jet identified as coming from the same top quark as the W boson~\cite{Mahlon_Parke_paper}.  Simulation studies show that this algorithm correctly identifies the down-type quark 60\% of the time.

The same-helicity and opposite-helicity model templates are created with a customized version of the {\sc herwig} event generation software package~\cite{hwgman} that implements the angular distribution of Eq.~\ref{decay_correlation} for the charged lepton or down-type quark, with a tunable choice of right- or left-handed top quarks, and preserves all the other expected spin correlations~\cite{genesThesis}.  We create four different simulated samples, corresponding to the four possible top-quark-pair helicity states: $\bar{t}_L t_{R}$, $\bar{t}_R t_{L}$, $\bar{t}_L t_{L}$, and $\bar{t}_R t_{R}$.  QCD interactions respect both the parity symmetry ($P$) and the combined symmetry of parity and charge conjugation ($CP$).  Because $CP$ transforms $\bar{t}_R t_R \rightarrow t_L \bar{t}_L$, we can define the same-helicity (SH) model template shape to be the symmetric sum of $\sigma(\bar{t}_R t_R) + \sigma(t_L \bar{t}_L)$.  Since $P$ transforms $\bar{t}_R t_L \rightarrow \bar{t}_L t_R$, we let the opposite-helicity (OH) model template shape be the symmetric sum of $\sigma(\bar{t}_R t_L) + \sigma(\bar{t}_L t_R)$.

\begin{figure}[!htbp]
\center
\includegraphics[width=0.49\textwidth]{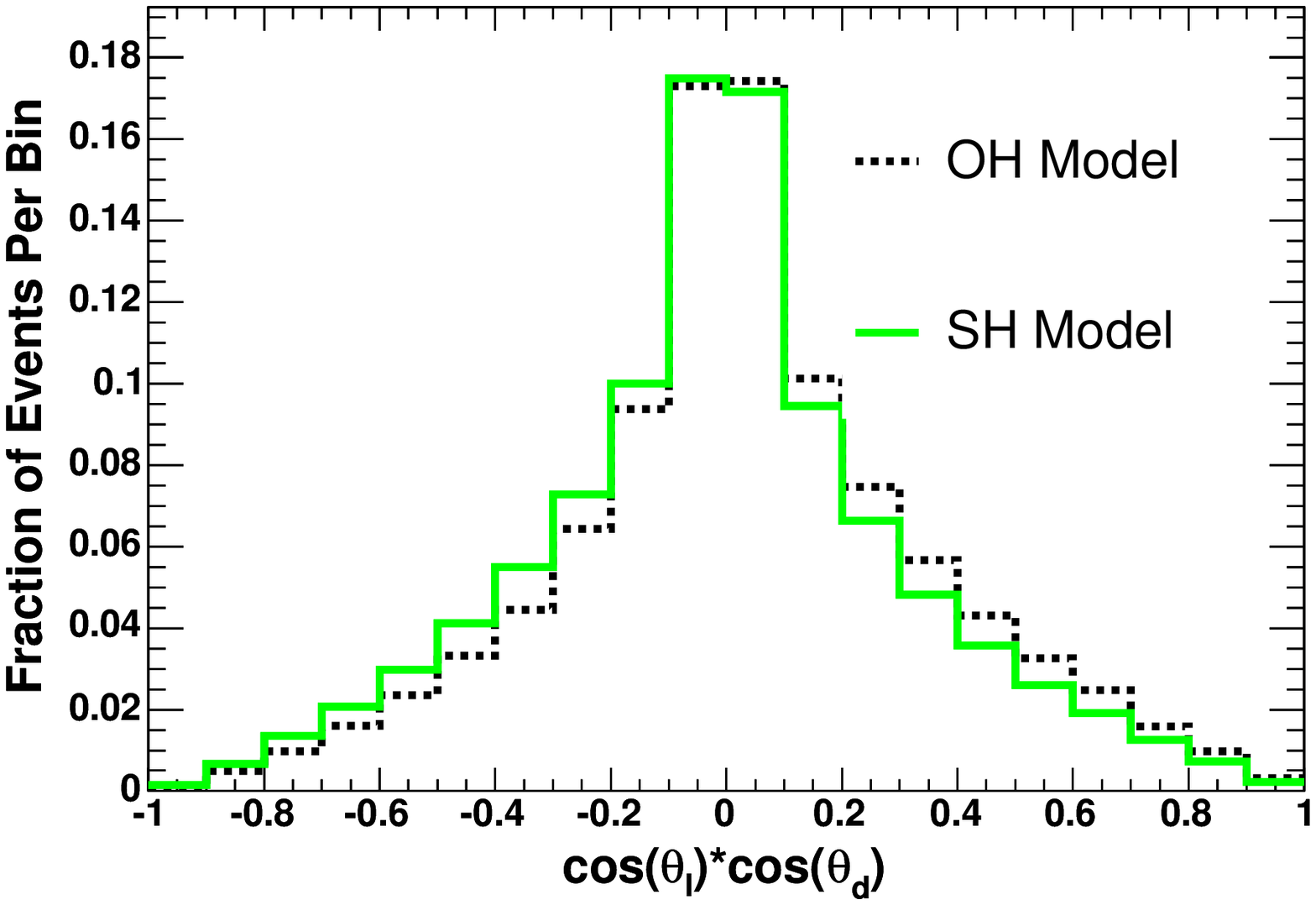}
\hfil
\includegraphics[width=0.49\textwidth]{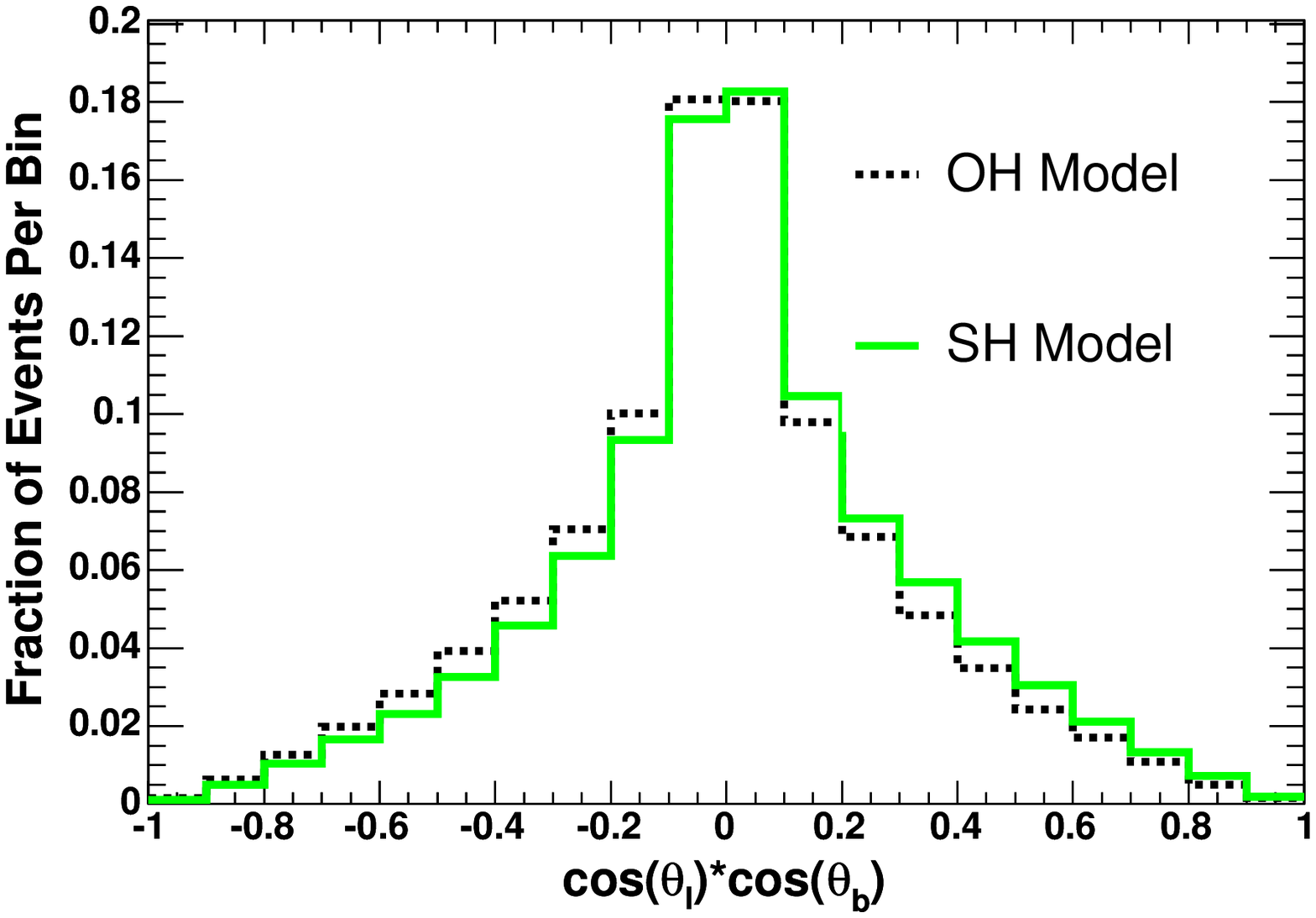}
\caption{Distributions of the $\cos\theta_l\cos\theta_d$ and $\cos\theta_l\cos\theta_b$ variables, after detector simulation, event selection, and reconstruction, in our same-helicity and opposite-helicity simulated $t\bar{t}$ samples. \label{alexei_comparison}}
\end{figure}

Figure~\ref{alexei_comparison} compares the SH and OH model templates after detector simulation, event selection, and reconstruction in the two distributions that we use for the measurement, $\cos\theta_l\cos\theta_d$ and $\cos\theta_l\cos\theta_b$.  Our sensitivity results from the SH model template being shifted towards negative values of $\cos\theta_l\cos\theta_d$, while the OH model template is shifted towards positive values, with the opposite shifts occurring in the $\cos\theta_l\cos\theta_b$ distribution.

We perform our measurement using a binned likelihood fit to find the relative normalization of these model templates that gives the best simultaneous representation of $\cos\theta_l\cos\theta_d$ and $\cos\theta_l\cos\theta_b$ in our data.  The background normalization is constrained to be close to the predicted value, with a Gaussian uncertainty, but the same-helicity fraction $F_{SH}$ and opposite-helicity fraction $F_{OH}$ are allowed to float freely.  We do not require that $F_{SH}$ and $F_{OH}$ be constrained to physical values between 0 and 1, but we do require $F_{SH} + F_{OH} = 1$. The fit runs over all bins in a two-dimensional distribution of $\cos\theta_l\cos\theta_d$ vs $\cos\theta_l\cos\theta_b$.  The expected statistical uncertainty for $F_{OH}$ is approximately 0.23, corresponding to an uncertainty for $\kappa$ of 0.46, and is independent of the actual value of $F_{OH}$ and $\kappa$.

\begin{table}[h]
\begin{center}
\caption{Systematic Uncertainties on $F_{OH}$} \label{syst}
\begin{tabular}{c c}
\hline
\hline
Systematic           & Uncertainty \\
\hline
Generator dependence    & $0.060$ \\
JES                     & $0.042$ \\
ISR/FSR                 & $0.030$ \\
Background shape        & $0.023$ \\
Color reconnection      & $0.009$ \\
PDF                     & $0.007$ \\
Parton shower           & $0.006$ \\
Background size         & $0.002$ \\
\hline
Total uncertainty       & $0.083$ \\
\hline
\hline
\end{tabular}
\end{center}
\end{table}

Additional contributions to the uncertainty result from incomplete knowledge of the background size and shape, of the exact detector response, and of the parton distribution functions (PDF), and are estimated by performing the measurement in simulated samples with reasonable variations in the model assumptions.  These systematic uncertainties are shown in Table~\ref{syst}.  The largest uncertainty, generator dependence, results from small biases seen when testing with simulated samples created using a range of event generation software packages, including {\sc herwig}~\cite{hwgman}, {\sc pythia}~\cite{pyman}, {\sc alpgen}~\cite{alpman}, and {\sc madevent}~\cite{mdeman}.  Other significant contributions come from the uncertainty of the jet energy scale (JES) during event reconstruction and uncertainty in the amount of initial and final state radiation (ISR/FSR) in our observed $t\bar{t}$ events.  The small variation of $F_{OH}$ with the assumed value of the top-quark mass is not included in our systematic uncertainty; our measurement assumes a mass of 172.5 GeV/$c^2$ for the top quark.

\begin{figure}[!htbp]
\center
\includegraphics[width=0.49\textwidth]{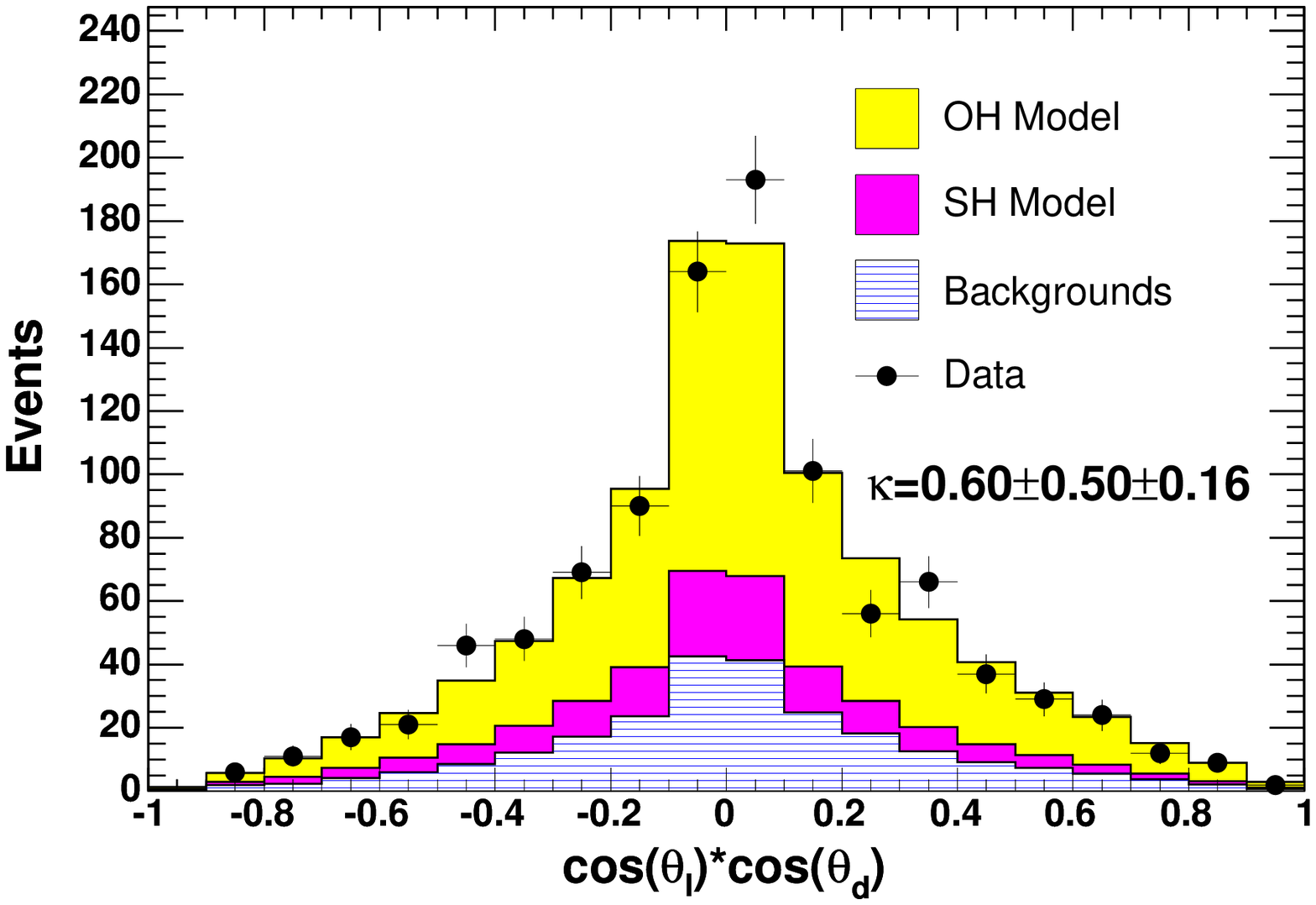}
\hfil
\includegraphics[width=0.49\textwidth]{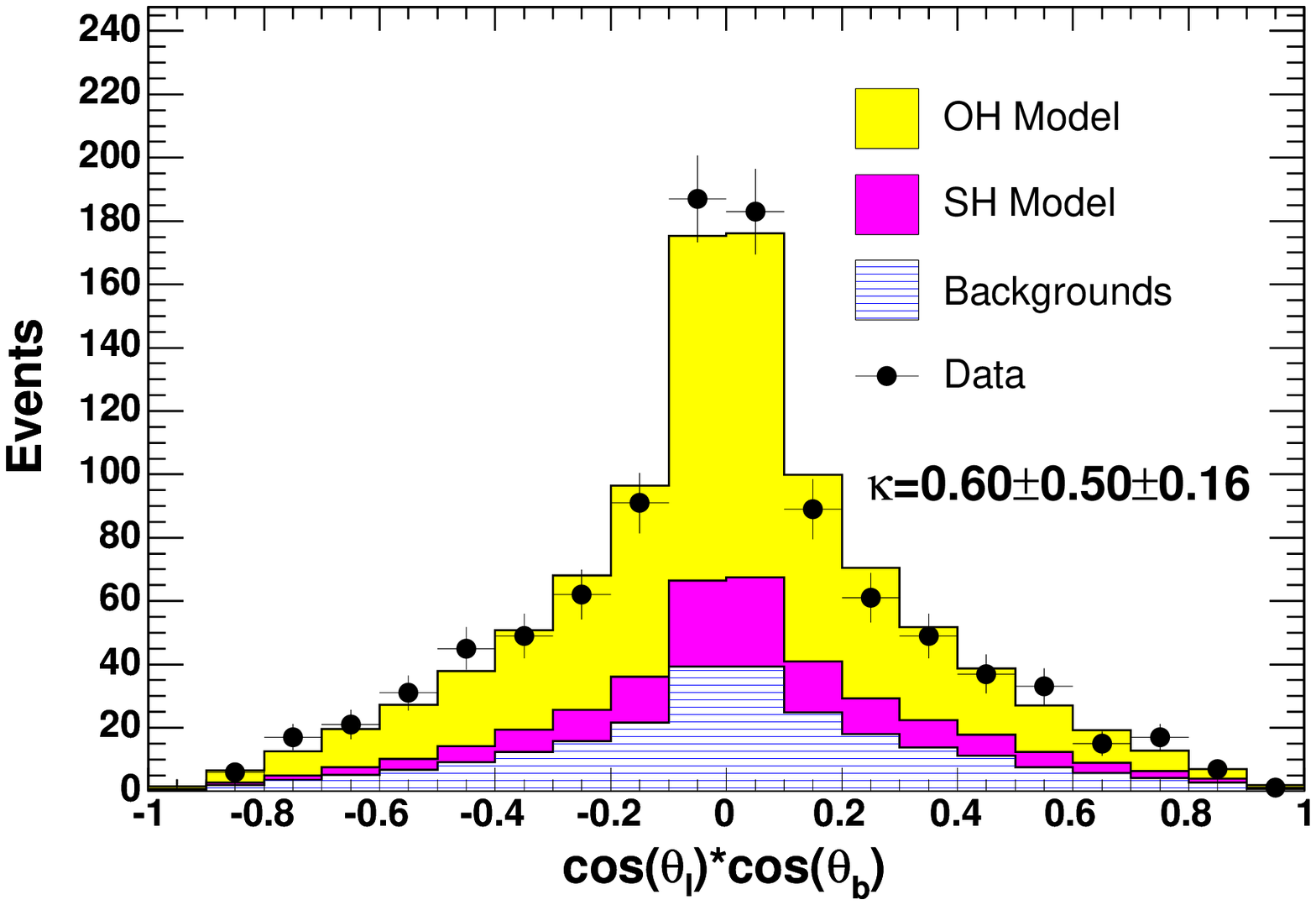}
\caption{Distribution of the $\cos\theta_l\cos\theta_d$ and $\cos\theta_l\cos\theta_b$ variables in data compared to the sum of our background model, the same-helicity model template, and the opposite-helicity model template. The relative normalizations of the model distributions are determined by our fit result.  \label{result_plots}}
\end{figure}

The final result of our fit to the two-dimensional distribution $\cos\theta_l\cos\theta_d$ vs. $\cos\theta_l\cos\theta_b$ is shown in Fig.~\ref{result_plots}.  This figure shows one-dimensional distributions of both variables, with our data being compared to the sum of the background model, same-helicity model, and opposite-helicity model, with the model normalizations determined by our fit result.  Assuming the top-quark mass is 172.5 GeV/$c^2$, we find an opposite-helicity fraction of

$$F_{OH}=0.80\pm0.25 \mathrm{\ (stat)}\pm0.08 \mathrm{\ (syst)}\ \ .$$

\noindent Converting this to the spin correlation coefficient, using $\kappa = 2F_{OH}-1$, yields

$$\kappa=0.60\pm0.50 \mathrm{\ (stat)}\pm0.16 \mathrm{\ (syst)}\ \ .$$

This first measurement of the top quark-pair spin correlation in the lepton plus jets decay channel agrees well with the theoretical prediction of $\kappa \approx 0.40$~\cite{willen,Mahlon_Parke_paper}, although the statistical uncertainty is still large.  Simulated experiments with larger datasets indicate that if the Tevatron dataset reaches 15 fb$^{-1}$ before the end of the Tevatron lifetime, the expected statistical uncertainty on $\kappa$ would be reduced to 0.26.  This technique can thus be applied in future measurements with larger datasets collected at the Tevatron and LHC to constrain the $t \bar{t}$ production spin structure or to connect with other anomalies that may show up in the reconstructable $t \bar{t}$ kinematics of the lepton plus jet sample.

We thank the Fermilab staff and the technical staffs of the participating institutions for their vital contributions. This work was supported by the U.S. Department of Energy and National Science Foundation; the Italian Istituto Nazionale di Fisica Nucleare; the Ministry of Education, Culture, Sports, Science and Technology of Japan; the Natural Sciences and Engineering Research Council of Canada; the National Science Council of the Republic of China; the Swiss National Science Foundation; the A.P. Sloan Foundation; the Bundesministerium f\"ur Bildung und Forschung, Germany; the World Class University Program, the National Research Foundation of Korea; the Science and Technology Facilities Council and the Royal Society, UK; the Institut National de Physique Nucleaire et Physique des Particules/CNRS; the Russian Foundation for Basic Research; the Ministerio de Ciencia e Innovaci\'{o}n, and Programa Consolider-Ingenio 2010, Spain; the Slovak R\&D Agency; and the Academy of Finland.


\begin{thebibliography}{9}
\bibitem{particledef} Throughout this Communication, when we refer to a $u$ quark, we mean either a $u$ or $c$ quark, and similarly, a $d$ quark refers to either a $d$ or $s$ quark.  Additionally, unless necessary, we do not differentiate between a particle and its antiparticle, so the inclusion of the charge-conjugate state should be assumed.
\bibitem{coords} In CDF's cylindrical coordinate system, the transverse momentum is defined as $p_T=p\sin\theta$, where $\theta$ is the polar angle measured from the proton direction (z-axis).  Similarly, the transverse energy is defined as $E_T=E\sin\theta$.  The missing transverse energy is defined by $\not\!\!E_T = |\not\!\!\vec{E}_T| = |- \sum_{i}E_T^i \hat{n}_i|$ where $\hat{n}_i$ is a unit vector perpendicular to the beam axis and pointing to the $i^{th}$ calorimeter tower.  The pseudorapidity is defined as $\eta = -\ln [\tan(\theta / 2)]$.
\end{thebibliography}
\end{document}